\documentclass[aps,prl,twocolumn,showpacs,superscriptaddress]{revtex4}
\usepackage{graphicx}

\begin{document}

\title{Three-dimensional Resonance in superconducting BaFe$_{1.9}$Ni$_{0.1}$As$_2$}
\author{Songxue Chi}
\affiliation{
Department of Physics and Astronomy, The University of Tennessee, Knoxville, Tennessee 37996-1200, USA
}
\author{Astrid Schneidewind}
\affiliation{Technische Universit\"{a}t Dresden,
 Institut f\"{u}r Festk\"{o}rperphysik, 01062 Dresden, Germany
}
\author{Jun Zhao}
\affiliation{
Department of Physics and Astronomy, The University of Tennessee, Knoxville, Tennessee 37996-1200, USA
}
\author{Leland W. Harriger}
\affiliation{
Department of Physics and Astronomy, The University of Tennessee, Knoxville, Tennessee 37996-1200, USA
}
\author{Linjun Li}
\affiliation{Department of Physics, Zhejiang University, Hangzhou 310027, China
}
\author{Yongkang Luo}
\affiliation{Department of Physics, Zhejiang University, Hangzhou 310027, China
}
\author{Guanghan Cao}
\affiliation{Department of Physics, Zhejiang University, Hangzhou 310027, China
}
\author{Zhu'an Xu}
\affiliation{Department of Physics, Zhejiang University, Hangzhou 310027, China
}
\author{Micheal Loewenhaupt }
\affiliation{Technische Universit\"{a}t Dresden,
 Institut f\"{u}r Festk\"{o}rperphysik, 01062 Dresden, Germany
}

\author{Jiangping Hu}
\affiliation{Department of Physics, Purdue University, West Lafayette, Indiana 47907, USA}
\author{Pengcheng Dai}
\email{daip@ornl.gov}
\affiliation{
Department of Physics and Astronomy, The University of Tennessee, Knoxville, Tennessee 37996-1200, USA
}
\affiliation{
Neutron Scattering Science Division, Oak Ridge National Laboratory, Oak Ridge, Tennessee 37831-6393, USA}

\begin{abstract}
We use inelastic neutron scattering to study magnetic
excitations of the FeAs-based superconductor
BaFe$_{1.9}$Ni$_{0.1}$As$_2$ above and below its superconducting
transition temperature $T_c=20$ K. In addition to gradually open a
spin gap at the in-plane antiferromagnetic ordering wavevector
$(1,0,0)$, the effect of superconductivity is to form a three
dimensional resonance with clear dispersion along the $c$-axis
direction. The intensity of the resonance develops like a
superconducting order parameter, and the mode occurs at
distinctively different energies at $(1,0,0)$ and $(1,0,1)$.  If
the resonance energy is directly associated with the
superconducting gap energy $\Delta$, then $\Delta$ is dependent on
the wavevector transfers along the $c$-axis. These results suggest
that one must be careful in interpreting the superconducting gap
energies obtained by surface sensitive probes such as scanning
tunneling microscopy and angle resolved photoemission.
\end{abstract}


\maketitle


Understanding the interplay between spin fluctuations and
superconductivity in high-transition-temperature (high-$T_c$)
superconductors is important because spin fluctuations may mediate
electron pairing for superconductivity \cite{scalapino,abanov}. In
the case of high-$T_c$ copper oxides, it is now well documented
that the spin fluctuation spectrum is dominated by a collective
excitation known as the resonance mode centered at the
antiferromagnetic (AF) ordering wavevector ${Q}=(1/2,1/2)$
\cite{rossat,mook,stock,fong99,wilson,ncco}. Although the
intensity of the mode behaves like an order parameter below $T_c$,
the energy of the mode is dispersionless for wavevector transfers
along the $c$-axis and directly tracks $T_c$
\cite{mook,stock,fong99,wilson,ncco}, thus suggesting that the
mode is an intrinsic property of the two-dimensional (2D) CuO$_2$
planes and intimately associated with superconductivity. For FeAs-based superconductors
\cite{kamihara,rotter,sefat,ljli}, the presence of static AF
ordering in their parent compounds (with spin structure 
of Fig. 1a) \cite{cruz,mcguire,jzhao1,qhuang,jzhao2,goldman} and
the remarkable similar doping dependent phase diagram to that of
the high-$T_c$ copper oxides \cite{jzhao1} suggest that AF spin
fluctuations may also play an important role in the
superconductivity of these materials. Indeed, recent neutron scattering measurements
on spin fluctuations of powder samples of superconducting
Ba$_{0.6}$K$_{0.4}$Fe$_2$As$_2$ ($T_c=38$ K) \cite{christianson}
and crystalline electric field (CEF) excitations of Ce in
CeFeAsO$_{0.84}$F$_{0.16}$ ($T_c=41$ K) \cite{chi} found clear
evidence for resonant-like magnetic intensity gain below
$T_c$ at $\hbar\omega\sim 14$ and 18.7 meV, respectively. However,
the Ce CEF measurements give no information on the $Q$-dependence of the scattering \cite{chi}. Although the
resonant-like scattering in Ba$_{0.6}$K$_{0.4}$Fe$_2$As$_2$ occurs
near the AF ordering wavevector, the powder nature of the
experiment impedes to distinguish whether the resonant scattering
is centered at the three-dimensional (3D) AF wavevector ${
Q}=(1,0,1)$ of its parent compound \cite{qhuang,jzhao2,goldman} or
simply at a 2D AF in-plane wavevector ${Q}=(1,0,0)$
\cite{christianson}.

In this Letter, we report the results of inelastic neutron
scattering studies of spin
fluctuations in single crystals of superconducting
BaFe$_{1.9}$Ni$_{0.1}$As$_2$ (with $T_c=20$ K, see Fig. 1c)
\cite{ljli}. We show that the effect of superconductivity is to gradually 
open a low-energy spin gap and also to induce a resonance at
energies above the spin gap energy.  Although the intensity of the
resonance develops below $T_c$ similar to that of the resonance in high-$T_c$ copper
oxides, the mode actually has a dispersion along the
$c$-axis, and occurs at distinctively different energies at ${Q}=(1,0,1)$ and $(1,0,0)$ in contrast with the cuprates. If the
resonance energy in FeAs superconductors is associated with $T_c$
and the superconducting gap energy $\Delta$, then $\Delta$ should
be 3D in nature and depend sensitively on the wavevector transfers
along the $c$-axis. Therefore, one must be careful
in interpreting $\Delta$ obtained by
surface sensitive probes such as scanning tunneling microscopy and
angle resolved photoemission.

We grew many high quality BaFe$_{1.9}$Ni$_{0.1}$As$_2$ single
crystals (each with mosaicity $<0.5^\circ$) using flux method as
described in Ref. \cite{ljli}. Figure 1c shows resistivity and
magnetic susceptibility data of a typical crystal showing an onset
$T_c$ of 20.2 K with a transition width less than 1 K.  We coaligned
21 single crystals on a flat Al plate to obtain a total mass of
about 0.6 grams.  The in-plane mosaic of the aligned crystal
assembly is about $1.3^\circ$ and the out-of-plane mosaic is less
than $4.3^\circ$ full width at half maximum (FWHM).
 Our neutron scattering experiments were performed on the PANDA cold triple-axis spectrometer
 at the Forschungsneutronenquelle Heinz Maier-Leibnitz (FRM II), TU M\"{u}nchen, Germany.
We used  pyrolytic graphite PG(0,0,2) as monochromator and
analyzer without any collimator. For inelastic neutron scattering, 
monochromator and analyzer were vertically and horizontally
focused. For sample alignment and elastic measurements they were
both horizontally flat and vertically focused.  We defined the
wave vector ${Q}$ at $(q_x,q_y,q_z)$ as
$(H,K,L)=(q_xa/2\pi,q_yb/2\pi,q_zc/2\pi)$ reciprocal lattice units
(rlu) using the orthorhombic magnetic unit cell previously taken for the magnetic structure determination
\cite{qhuang,jzhao2,goldman} and the low energy spin wave
measurements \cite{jzhao3,mcqueeney,matan} of the parent undoped
compound (space group $Fmmm$, $a=5.564$, $b=5.564$, and
$c=12.77$~\AA ).  We choose this reciprocal space notation
(although the actual crystal structure is tetragonal) for easy
comparison with previous spin-wave and elastic measurements on the
parent compound, where magnetic Bragg peaks and low-energy spin
waves are expected to occur around $(1,0,1)$ and $(1,0,3)$ rlu
positions (see Fig. 1b). For the experiment, the
BaFe$_{1.9}$Ni$_{0.1}$As$_2$ crystal assembly was mounted in the
$[H,0,L]$ zone inside a closed cycle refrigerator. The final
neutron wavevector was fixed at either $k_f=1.55$~\AA$^{-1}$ with
a cold Be filter or at $k_f=2.662$~\AA$^{-1}$ with a PG filter in
front of the analyzer.

\begin{figure}[t]
\includegraphics[scale=.4]{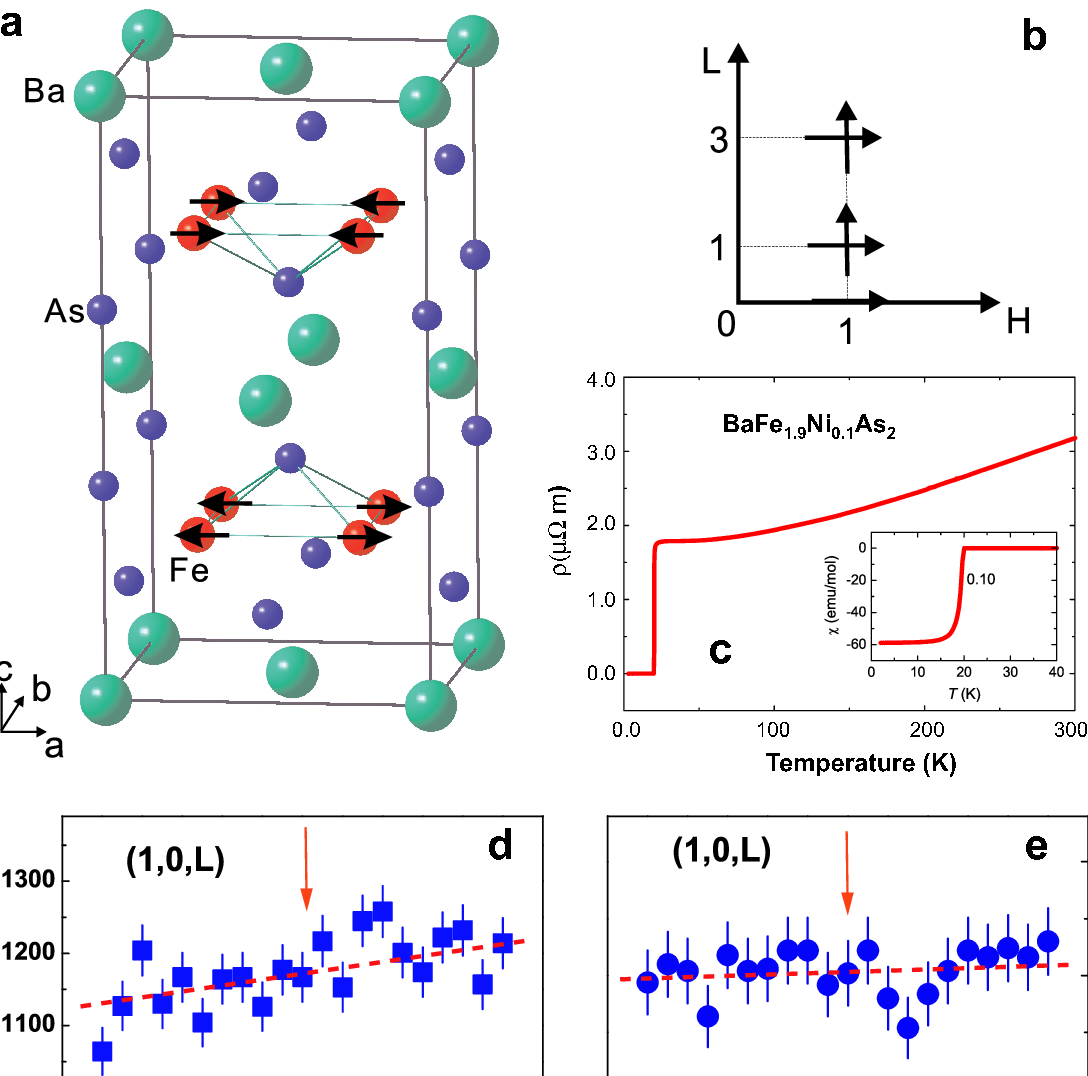}
\caption{(color online). (a) Schematic diagram of the Fe spin ordering in the BaFe$_2$As$_2$ and we use the same unit cell for 
BaFe$_{1.9}$Ni$_{0.1}$As$_2$ for easy comparison.
(b) Reciprocal space probed in our experiment. (c) Resistivity and 
 Magnetic susceptibility measurements of $T_c$. 
(d,e) Elastic neutron scattering $L$-scans through $(1,0,1)$ and $(1,0,3)$ magnetic Bragg peaks at 30 K, showing no evidence of
static long-range AF order \cite{jzhao2,goldman}.  
 }
\end{figure}

\begin{figure}[t]
\includegraphics[scale=.4]{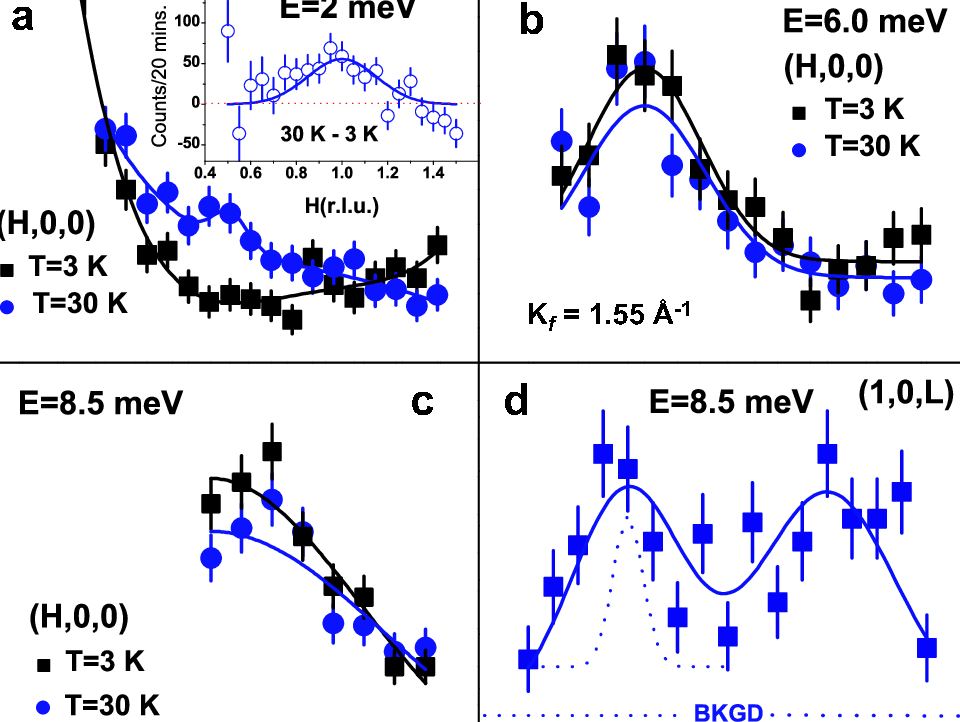}
\caption{(color online). Constant energy scans around the $(1,0,0)$ and $(1,0,1)$ positions 
for $\hbar\omega=2$, 6, and 8.5 meV obtained with $k_f=1.55$~\AA$^{-1}$.
(a-c) $Q$-scan along the $[H,0,0]$ direction for $\hbar\omega=2$, 6, and 8 meV at 30 K and 3 K.  The inset
in (a) shows the temperature difference plot and a Gaussian fit to the data.
The missing low-$Q$ data for scans in (b) and (c) are due to kinematic constraint.
(d) $Q$-scan along the $[1,0,L]$ direction for $\hbar\omega=8.5$ meV at 3 K. Note two clear peaks centered at 
$(1,0,-1)$ and $(1,0,1)$, respectively. The dashed-line peak is the low-temperature 
spin-waves of BaFe$_2$As$_2$ at $\hbar\omega=10$ meV from Fig. 2f in \cite{matan}. 
 }
\end{figure}

We first searched for possible static AF order in our samples. For
undoped BaFe$_2$As$_2$, magnetic Bragg peaks are expected at the
$(1,0,1)$ and $(1,0,3)$ positions for the spin structure of Fig.
1a \cite{qhuang}. In addition, the low temperature spin waves are
gapped below 9.8~meV \cite{matan}.
Our elastic $Q$-scans through these expected AF Bragg peak
positions are featureless (Figs. 1d and 1e), confirming the
absence of static long-range order above 30~K. Figure 2 summarizes
constant-energy scans at 3~K (well below $T_c$) and at 30~K (above
$T_c$) at $\hbar\omega=2$, 6, and 8.5~meV carried out with
$k_f=1.55$~\AA$^{-1}$.  Although these probed energies are well below the
9.8~meV spin gap energy in the parent compound \cite{matan}, we
observe at 30~K clear peaks centered at the in-plane AF wavevector
$(1,0,0)$ for $\hbar\omega=2$ and 6~meV, and half of a peak at
$\hbar\omega=8.5$~meV due to kinematic constraints (Figs. 2a-c).
Fourier transforms of the gaussian peaks in Figs. 2a and 2b gave the minimum 
dynamic spin correlation lengths of $\xi\approx 16\pm4$ and $21\pm4$ \AA\
for $\hbar\omega=2$ and 6~meV, respectively. 
The spin-spin correlations extend only to several chemical unit cells and are 
much smaller than the $\xi\approx 80\pm10$ \AA\ 
at $\hbar\omega=1.5$ meV obtained for electron-doped cuprate superconductor 
Pr$_{0.88}$LaCe$_{0.12}$CuO$_4$ \cite{wilson}.
On cooling from the
normal ($T=30$~K) to the superconducting ($T=3$~K)
state, the Gaussian peak at $\hbar\omega=2$~meV vanishes and
suggests the opening of a spin gap (Figs. 2a). In contrast, the
Gaussian peaks at $\hbar\omega=6$~meV hardly change across $T_c$
(Fig. 2b) whereas the scattering at $(1,0,0)$ for
$\hbar\omega=8.5$~meV actually increases below $T_c$ (Fig. 2c).
These results are similar to those for electron-doped
Nd$_{1.85}$Ce$_{0.15}$CuO$_4$ \cite{ncco}, and immediately suggest
that the opening of a low-energy spin gap below $T_c$ is
compensated by intensity gain above the gap energy. The low
temperature $(1,0,L)$ scan at $\hbar\omega=8.5$~meV shows two
broad peaks centered at $(1,0,-1)$ and $(1,0,1)$ corresponding to
the 3D AF ordering wavevector \cite{qhuang,jzhao2,goldman}.

\begin{figure}[t]
\includegraphics[scale=.40]{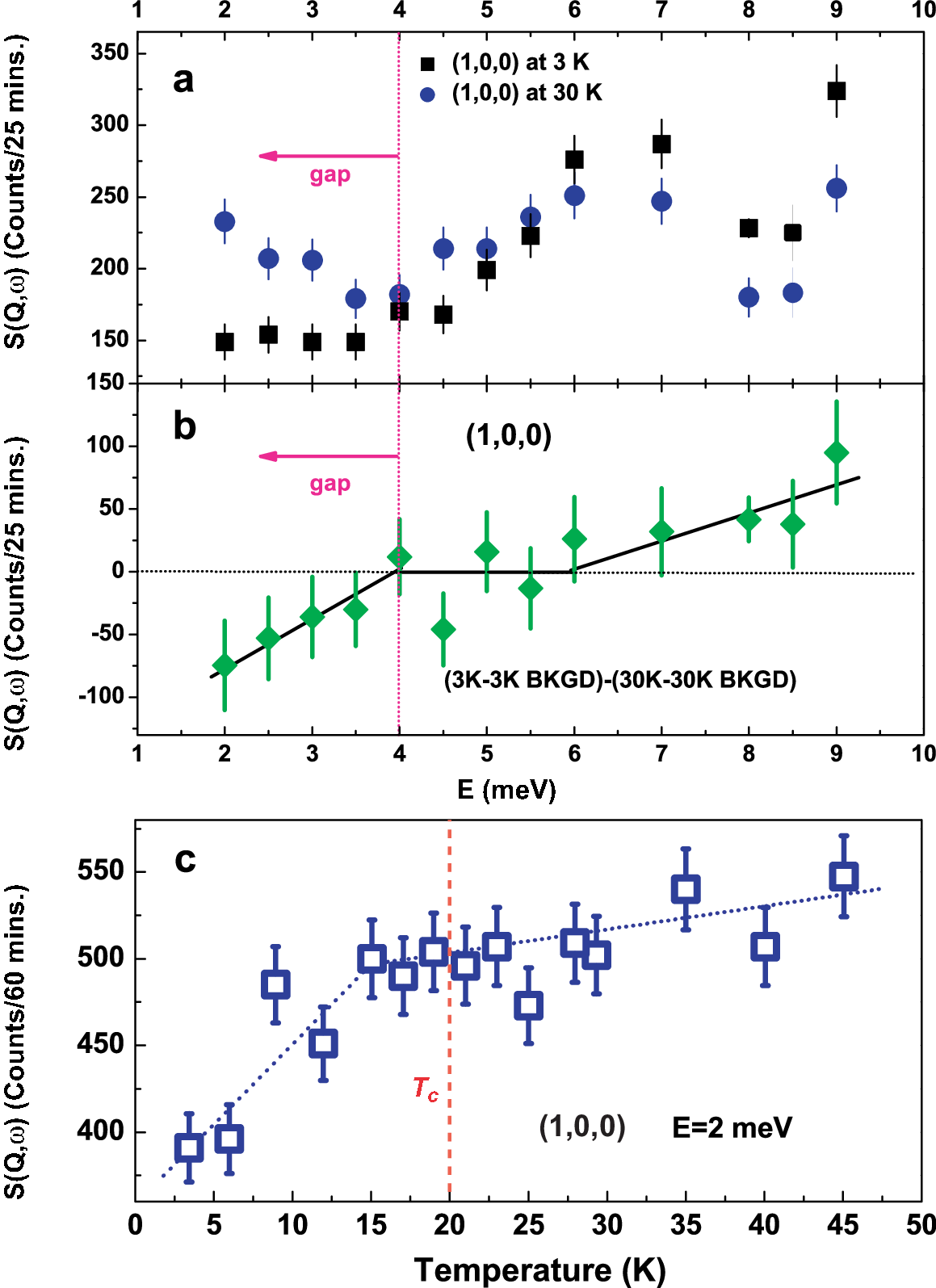}
\caption{(color online). Constant-$Q$ scans at the $(1,0,0)$ position 
above and below $T_c$ obtained with $k_f=1.55$~\AA$^{-1}$ and the temperature dependence of the scattering
at $Q=(1,0,0)$ and $\hbar\omega=2$ meV.  Background scattering was also collected at the $(1.3,0,0)$ position. 
(a) Energy scans at $Q=(1,0,0)$ from 2 meV to 9 meV at 30 K and 3 K.  (b) Background corrected intensity difference between
the 3 K and 30 K data at $Q=(1,0,0)$.  The negative scattering below 4 meV indicates 
the opening of a spin gap, while positive scattering above 6 meV suggests magnetic intensity gain below $T_c$. 
(c) Temperature dependence of the scattering obtained at $Q=(1,0,0)$ and $\hbar\omega=2$ meV with vertical dashed line indicating
the onset of superconductivity.  The $\hbar\omega=2$ meV spin gap does not start to open until about 4 K below $T_c$.
 }
\end{figure}

\begin{figure}[t]
\includegraphics[scale=.4]{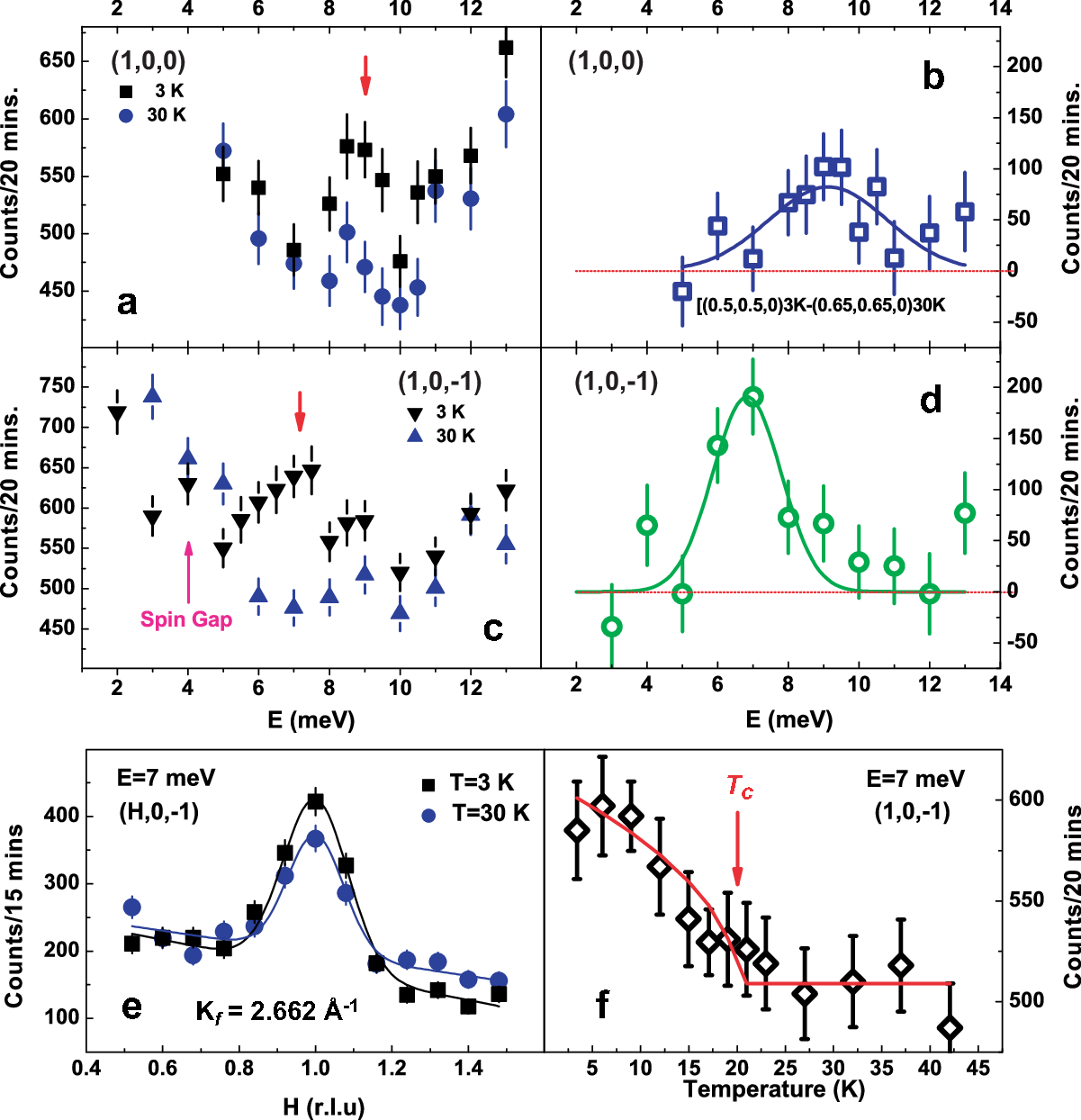}
\caption{(color online). Constant energy and constant-$Q$ scans around the $Q=(1,0,0)$ and $(1,0,1)$ positions 
above and below $T_c$ obtained with $k_f=2.662$~\AA$^{-1}$. We also collected background scattering at $Q=(1.3,0,-1)$ and found it to be temperature
independent between 30 K and 3 K for energies above 5 meV (not shown).
(a) Energy scans at the 2D AF wavevector $Q=(1,0,0)$ from 5 meV to 13 meV at 30 K and 3 K. Clear magnetic intensity gain is observed at 
$\hbar\omega=9.0$ meV. (b) 
The temperature difference scattering between 3 K and 30 K shows a clear resonant peak at $\hbar\omega=9.1\pm0.4$ meV.  
(c) Energy scans at the 3D AF wavevector $Q=(1,0,-1)$ from 2 meV to 13 meV at 30 K and 3 K. The resonance has shifted to 
 $\hbar\omega=7$ meV and is about 2.0 meV lower than that at $Q=(1,0,0)$. (d) The temperature difference plot confirms that the mode has now moved to
 $7.0\pm0.5$ meV. Note that the intensity gain below $T_c$ at $\hbar\omega=7$ meV and $Q=(1,0,-1)$ is just as large as that at 
 $\hbar\omega=9.1$ meV and $Q=(1,0,0)$.
 (e) Wavevector dependence of the scattering at 30 K and 3 K for $\hbar\omega=7$ meV, confirming that the resonance intensity gain occurs at 
 $Q=(1,0,-1)$.
(f) Temperature dependence of the scattering obtained at $Q=(1,0,-1)$ and $\hbar\omega=7$ meV shows clear order parameter like increase below $T_c$.
 }
\end{figure}

To determine the size of the superconducting spin gap and confirm
the intensity gain at $\hbar\omega=8.5$~meV below $T_c$, we
carried out energy scans at the in-plane AF wavevector $Q=(1,0,0)$ below and above $T_c$
(Fig. 3a). While the background scattering 
collected at $Q=(1.3,0,0)$ (not shown) changes only negligibly
between 30~K and 3 ~K, intensity at $Q=(1,0,0)$ is suppressed for
$\hbar\omega\leq 4$~meV and enhanced for $\hbar\omega\geq 6$~meV
with highest intensity at $\hbar\omega=9$~meV. In Fig. 3b we plot
the difference (3~K minus 30~K) of the (background corrected)
magnetic scattering at $Q=(1,0,0)$, again confirming the opening
of a spin gap for $\hbar\omega\leq 4$~meV and enhanced magnetic
scattering for $\hbar\omega\geq 6$~meV in the superconducting
state. Figure 3c shows the temperature dependence of the
scattering at $Q=(1,0,0)$ and $\hbar\omega=2$~meV. It seems that
the suppression of intensity at $\hbar\omega=2$~meV does not
exactly start at $T_c$ but about 4~K below $T_c$. These results
suggest that the spin gap in BaFe$_{1.9}$Ni$_{0.1}$As$_2$ opens
gradually with decreasing temperature until it reaches about 4~meV
at 3~K, remarkable similar to the spin gap behavior of
electron-doped Nd$_{1.85}$Ce$_{0.15}$CuO$_4$ \cite{ncco,yamada}.

Although the results displayed in Figs. 1-3  using
$k_f=1.55$~\AA$^{-1}$\ are suggestive of a resonance below $T_c$,
kinematic constraints did not allow us to carry out measurements
for energies above $\hbar\omega=9$~meV at $Q=(1,0,0)$.  To
determine the energy location of the possible mode, we collected
additional data with $k_f=2.662$~\AA$^{-1}$. Figure 4a shows the energy scan 
raw data at $Q=(1,0,0)$ below and above $T_c$.  Inspection of the data reveals that 
the low-temperature scattering enhances dramatically around
$\hbar\omega=9.0$~meV compared to the normal state scattering.
Since Bose population factor does not contribute much to magnetic scattering intensity 
for $\hbar\omega\geq5$ meV between 3 K and 30 K, 
the (low minus high) temperature difference scattering represents the 
net magnetic intensity gain at low temperature.
Subtracting the 30~K data from the~3 K data reveals a clear
localized mode near 9.0 meV (Fig. 4b).  Gaussian fit to the data gives a peak position 
$\hbar\omega=9.1\pm0.4$~meV, a peak width $3.3\pm0.9$ meV, and an integrated area 
$346\pm82$ per 20 minutes (Fig. 4b).

Since spin excitations at 
$\hbar\omega=8.5$~meV peak at $(1,0,-1)$/$(1,0,1)$ and are clearly dispersive along the
$c$-axis direction (Fig. 2d), we carried out additional measurements to search for resonance 
at the 3D AF ordering wavevector $Q=(1,0,-1)$ below and above $T_c$.  
The outcome in Fig. 4c shows a large magnetic intensity gain below $T_c$ at
$\hbar\omega=7$ meV, clearly different from the 9.1 meV resonance at $Q=(1,0,0)$.  
A Gaussian fit to the temperature difference plot in Fig. 4d gives a peak position
$\hbar\omega=7.0\pm0.5$ meV, a peak with $1.9\pm0.7$ meV, and an integrated area of $464\pm 145$ per 20 minutes.
To further confirm that the intensity gain at $\hbar\omega=7$ meV is indeed the resonance occurring at $Q=(1,0,-1)$, we carried constant-energy scans around $(1,0,-1)$ and outcome clearly shows that the intensity gain below $T_c$ 
arises from scattering at the 3D AF ordering position (Fig. 4e).  Finally, in
Fig. 4f we plot the temperature dependence of the scattering at $(1,0,-1)$ 
and $\hbar\omega=7$ meV.  The scattering increases dramatically below the
onset of $T_c$ and is remarkably similar to that of the resonance in high-$T_c$ copper oxides \cite{rossat,mook,stock,fong99,wilson,ncco}.

If the resonance is a measure of electron pairing correlations in high-$T_c$ superconductors \cite{dai}, the observed 3D resonance dispersion 
in BaFe$_{1.9}$Ni$_{0.1}$As$_2$ would suggest a variation of the superconducting energy gap $\Delta$ along the $c$-axis.
This is quite different from the high-$T_c$ copper oxides, where $\Delta$ is strictly 2D and independent of the $c$-axis modulations. For FeAs-based superconductors, 
the resonance may arise from quasiparticle transitions across the 
sign-revised $s$-wave electron ($\Delta^e_0$) and hole ($\Delta^h_0$) superconducting gaps 
in pure two dimensional models \cite{mazin08,maier,korshunov,chubukov,seo08,wang08}. 
By considering the AF coupling between layers,  the gap
functions can be naturally modified
to $\Delta_e(k_z)=\Delta^0_e+\delta \cos(k_z)$ and
$\Delta_h(k_z)=\Delta^0_h+\delta \cos(k_z)$. For a sign-revised $s$ pairing
symmetry, $\Delta^e_0 \sim -\Delta^h_0\sim -\Delta_0$. Therefore, the
dispersion of the resonance along $c$-axis is roughly determined
by \cite{maier}
\begin{eqnarray}
\hbar\omega(q_z) & \sim & Min(<|\Delta_e(k_z)|+
|\Delta_h(k_z+q_z)|>, k_z)\nonumber\\ &\sim& 2\Delta_0-2\delta
|\sin(\frac{q_z}{2})|
\end{eqnarray}
Based on this interpretation, our experimental results suggest
$\delta/{\Delta_0} =
[\omega(1,0,0)-\omega(1,0,-1)]/\omega(1,0,0)=0.26\pm0.07 $. If spin fluctuations are responsible for electron pairing and superconductivity, the values $\Delta_0$ and $\delta$ are expected
to be proportional to the intra-plane and inter-plane AF couplings, $J_{\parallel}$ and $J_{\perp}$,
respectively, which naturally suggests ${\delta}/{\Delta_0}\sim
{J_{\perp}}/{J_{\parallel}}$. The ratio
${\delta}/{\Delta_0}$ determined by our resonance dispersion is
reasonable agreement with the ratio of the AF exchange
couplings measured by neutron scattering experiments in the parent
compounds\cite{jzhao3,mcqueeney,matan}. These results suggest that  
spin fluctuations are also important for superconductivity in 
FeAs-based superconductors.

This work is supported by the U.S. NSF No. DMR-0756568, U.S. DOE BES No.
DE-FG02-05ER46202, and in part by the U.S. DOE, Division of Scientific
User Facilities. The work at Zhejiang University is supported by the NSF of China.
We further acknowledge support from DFG
within Sonderforschungsbereich 463 and from the PANDA project of
TU Dresden and FRM II.

Note added:  After independently finishing the experimental part of the present paper, 
we became aware of a similar neutron scattering experiment, where the resonance at $\hbar\omega=9.5$ meV 
was discovered near $Q=(1,0,0)$ [$(1/2,1/2,0)$ in tetragonal notation] 
in single crystal superconducting 
BaFe$_{1.84}$Co$_{0.16}$As$_2$ ($T_c=22$ K) \cite{lumsden}.  


\end{document}